\pdfoutput=1

\documentclass[11pt]{article} % use larger type; default would be 10pt

\usepackage[utf8]{inputenc} % set input encoding (not needed with XeLaTeX)
\usepackage{geometry} % to change the page dimensions
\usepackage{graphicx} % support the \includegraphics command and options
\usepackage{booktabs} % for much better looking tables
\usepackage{array} % for better arrays (eg matrices) in maths
\usepackage{paralist} % very flexible & customisable lists (eg. enumerate/itemize, etc.)
\usepackage{verbatim} % adds environment for commenting out blocks of text & for better verbatim
\usepackage{subfig} % make it possible to include more than one captioned figure/table in a single float
\usepackage{fancyhdr} % This should be set AFTER setting up the page geometry
\usepackage{sectsty}
\usepackage[nottoc,notlof,notlot]{tocbibind} % Put the bibliography in the ToC
\usepackage[titles,subfigure]{tocloft} % Alter the style of the Table of Contents

\usepackage{amsmath}
\usepackage{amssymb}
\usepackage{amsthm}
\usepackage[usenames, dvipsnames]{color}

\allsectionsfont{\sffamily\mdseries\upshape} % (See the fntguide.pdf for font help)
\title{When is $Y_{obs}$ missing and $Y_{mis}$ observed?}
\author{JC Galati$^1$}
\date{%
    \today \\
    \mbox{}\\%
    $^1$\small{Department of Mathematics and Statistics, La Trobe University, Melbourne, VIC 3083}
}

\begin{document}
\maketitle

\begin{abstract}
   In statistical modelling of incomplete data, missingness is encoded as a relation between datasets $Y$ and
   missingness patterns~$R$. The partitioning of $Y$ into observed and missing components is often denoted
   $Y_{obs}$ and~$Y_{mis}$. We point out a mathematical defect in this notation which results from two different
   mathematical relationships between $Y$ and $R$ not being distinguished, $(Y_{obs}, Y_{mis}, R)$ in which
   $Y_{obs}$ values are always observed, and $Y_{mis}$ values are always missing, and the overlaying of a
   missingness pattern onto the marginal distribution for~$Y$, denoted~$(Y_{obs}, Y_{mis})$. With the latter,
   $Y_{obs}$ and $Y_{mis}$ each denote mixtures of observable and unobservable data. This overlaying of the
   missingness pattern onto $Y$ creates a link between the mathematics and the meta-mathematics which violates the
   stochastic relationship encoded in $(Y, R)$. {Additionally, i}n the theory there is a need to compare partitions of $Y$
   according to different missingness patterns simultaneously. A simple remedy for these problems is to use four symbols
   instead of two, and to make the dependence on the missingness pattern explicit. We explain these and related issues.

   \vspace*{3mm}
   \noindent
   \textit{Key words and phrases:}
   incomplete data, missing data, ignorable, ignorability, missing at random, multiple imputation.
\end{abstract}

% ==========================================================================
\section{Introduction} \label{Sect:Introduction}
% ==========================================================================
The modern framework for statistical modelling of incomplete data was introduced by Rubin\;(1976).
Alongside the vector of data random variables,~$Y$, a vector of {response} binary random variables~$R$ was
introduced, and conditions were given under which inferences could be based on the marginal
density~$f(\mathbf{y})$ alone. Note that $U$ and~$M$ were used in~Rubin\;(1976) to denote what we have called
$Y$ and~$R$, respectively.

Intrinsic to this approach is the partitioning of a realisation $\mathbf{y}$ of $Y$ into values that are observed and
values that are missing according to some {missingness} pattern~$\mathbf{r}$. In Rubin (1976) the subscripts `$(1)$'
and `$(0)$' were introduced to denote this partition. These were replaced with the subscripts in $Y_\text{obs}$ and
$Y_\text{nob}$ in Rubin\;(1987) and $Y_\text{obs}$ and $Y_\text{mis}$ in~Little and Rubin\;(1987) and
Schafer\;(1997). Over three decades the latter notation has become a de\,facto standard in the exposition of statistical
methods for incomplete data typically aimed at practicing statisticians and other investigators, so it is important for there
to be a clear understanding of what it means.

% ==========================================================================
\section{There are two different relationships between $Y$ and~$R$} \label{Sect:TwoNotionsOfMissingness}
% ==========================================================================
The following is an extract from Little and Rubin\;(1987\;pp\;89--90); also see Little and Rubin\;(2002,\;pp\;118-119):

\vspace*{2mm}
\textit{
``Here to keep the notation simple we will be somewhat imprecise in our treatment of these complications. ...
}

\noindent
\textit{
\newline
\quad ...
The actual observed data consists of the values of the variables $(Y_{obs}, R)$. The distribution of the observed
data is obtained by integrating $Y_{mis}$ out of the joint density of $Y=(Y_{obs}, Y_{mis})$ and~$R$. That is,
\begin{equation}
   \mspace{83mu}
   f(Y_{obs}, R|\theta, \psi) \;=\;
      \int
      f(Y_{obs}, Y_{mis}|\theta)\,
      f(R|Y_{obs}, Y_{mis}, \psi)\,
      \text{d}Y_{mis}.
   \notag \mspace{35mu} (5.11)\text{''}
\end{equation}
}

In the extract above, the authors stated that their intention was to keep the notation simple. But setting
$Y = (Y_{obs}$, $Y_{mis})$ encodes missingness into the notation as attributes of the data vector~$Y$ instead of as
the vector~$(Y_{obs}$, $Y_{mis}, R)$ in the mathematical relation $(Y, R)$. This, in fact, significantly complicates
rather than simplifies the notation, particularly in regard to the domain of the marginal density, $f(Y|\theta)$,
for~$Y$. In the product of functions $f(-|\theta)f(-|-,\psi)$ on the right hand side of~(5.11), the factor
$f(-|\theta)$ is shorthand for the composition of functions $f\circ\pi_Y$, where $\pi_Y$ is the projection sending a
realisation~$(\mathbf{y}, \mathbf{r})$ of $(Y, R)$ to the realisation $\mathbf{y}$ of~$Y$. When unpacked this
way, the notation in (5.11) specifies that
\begin{equation}
   (Y_{obs}, Y_{mis}) \,=\, \pi_Y(Y_{obs}, Y_{mis}, R).
   \label{Eq:Projection}
\end{equation}
This is not straightforward to interpret because the formal mathematical relation of missingness exists in the domain
of~$\pi_Y$, and this missingness relation is not preserved by the projection~$\pi_Y$.

Note that the failure of~$\pi_Y$ to preserve the missingness relation is not simply because $\pi_Y$ is a many-to-one
function. Even if the domain of $\pi_Y$ is restricted to include only pairs pertaining to a specific {missingness}
pattern~$\mathbf{r}$, the two missingness relationships still differ. The formal definition of `observed' and `missing'
encoded in $(Y, R)$ on the right hand side of~(\ref{Eq:Projection}) is an absolute concept: every data item in the
range of~$(Y, R)$ is stamped irrevocably either as `observed' or `missing'. On the left hand side, however, 
`observed' and `missing' mean `observed this time' and `missing this time', respectively. This is a different concept
which at the meta-mathematical level is inconsistent with~$(Y, R)$ {at the stochastic level(delete)}: in the density functions
$f(\mathbf{y}_{obs}, \mathbf{y}_{mis})$ and~$f(\mathbf{y}_{mis} |\, \mathbf{y}_{obs})$, the
notations~$\mathbf{y}_{obs}$ and~$\mathbf{y}_{mis}$ denote arbitrary realisations, which entails holding fixed
the {missingness} pattern determining the partition of~$Y$ while allowing
$\mathbf{y}=(\mathbf{y}_{obs}, \mathbf{y}_{mis})$ in the marginal distribution to vary (in contradiction of the
stochastic relationship encoded in~$(Y, R)$).

To distinguish these different concepts, the mathematical relationship $(Y_{obs}, Y_{mis}, R)$ will be called
\textbf{formally missing} and the relationship $(Y_{obs}, Y_{mis})$ \textbf{temporally missing}. We note that there
is an additional need in the theory to extend the temporal missingness relationship to all of  $Y\times R$ by partitioning
the components of $Y$ in all of $Y\times R$ according to some fixed {missingness} pattern $\mathbf{r}$ (that is, over all
the possible {missingness} patterns). The use of this will be illustrated in the derivations in Appendix\;C and Appendix\;E.

Informally, the distinction between formal and temporal missingness is that the former is what is defined formally
by the relation~$(Y, R)$, whereas with the latter the data variables $Y$ have been partitioned according to some
{missingness} pattern~$\mathbf{r}$ simply for the purpose of considering $Y$ from a particular point of view, and
there is no requirement or expectation that the formal relationship $(Y, R)$~is, or can be, preserved.

Note that when formal missingness is intended, $Y_{obs}$ makes sense only as one part of a pair~$(Y_{obs}, R)$,
and this pair denotes a stochastic function more general than a random vector. On the other hand, with temporal
missingness, both $Y_{obs}$ and $Y_{mis}$ denote marginal distributions of $Y$ that are each mixtures of
formally observable and formally unobservable values.

% ==========================================================================
\section{There are also two different functions $f(\mathbf{y}_{mis} |\, \mathbf{y}_{obs})$}
   \label{Sect:TwoDifferentFunctions}
% ==========================================================================
\begin{center}
   (For this and subsequent sections see Appendix\;A for definitions of notation.)
\end{center}
A statement that is equivalent to a missing at random (MAR) assumption is often written in the following  (or a similar)
way:
\begin{equation}
   p(\mathbf{y}_{mis} |\, \mathbf{y}_{obs}, \mathbf{r}) \,=\,
      f(\mathbf{y}_{mis} |\, \mathbf{y}_{obs}).
   \label{Eq:WrongMARIdentity}
\end{equation}
Sometimes (\ref{Eq:WrongMARIdentity}) is assumed to hold for just the realised observed
values~$(\widetilde{\mathbf{y}}_{obs}, \widetilde{\mathbf{r}})$ and at other times it is assumed to hold for all
possible observable values $(\mathbf{y}_{obs}, \mathbf{r})$ under repeated sampling from~$(Y, R)$
(see Seaman\;et.\,al.\,(2013) for details).

Despite the function on the right hand side of~(\ref{Eq:WrongMARIdentity}) being denoted 
`$f(\mathbf{y}_{mis} |\, \mathbf{y}_{obs})$',
technically the functions being compared are $p(\mathbf{y}_{mis} |\, \mathbf{y}_{obs}, \mathbf{r})$ on
the left hand side and
$f(-|-)\circ\pi_Y|_{\Omega_{\mathbf{r}}}(\mathbf{y}_{mis}, \mathbf{y}_{obs}, \mathbf{r})$
on the right hand side, where $\pi_Y|_{\Omega_{\mathbf{r}}}$ is the restriction of the projection $\pi_Y$
to the domain of $p(\mathbf{y}_{mis} |\, \mathbf{y}_{obs}, \mathbf{r})$ and $f(-|-)$ denotes the
function $f(\mathbf{y}_{mis} |\, \mathbf{y}_{obs})$ derived from the marginal density
$f(\mathbf{y}_{mis}, \mathbf{y}_{obs})$ for~$Y$.
Note that
$f(-|-)\circ\pi_Y|_{\Omega_{\mathbf{r}}}(\mathbf{y}_{mis}, \mathbf{y}_{obs}, \mathbf{r})
   \neq f(\mathbf{y}_{mis} |\, \mathbf{y}_{obs})$
because these functions have different domains.
This mathematical distinction is a minor technicality, but the distinction is important stochastically.
We will illustrate this shortly, but first we distinguish between these two functions by giving them different notation:
\begin{alignat}{1}
   f^{(T)}(\mathbf{y}_{mis} |\, \mathbf{y}_{obs}) \,&:=\,
      f(\mathbf{y}_{mis} |\, \mathbf{y}_{obs})
      \label{Eq:Wrongf} \\
   f^{(F)}(\mathbf{y}_{mis} |\, \mathbf{y}_{obs}) \,&:=\,
      f(-|-)\left(\,\pi_Y(\mathbf{y}_{mis}, \mathbf{y}_{obs}, \mathbf{r})\,\right).
   \label{Eq:Rightf}
\end{alignat}

The stochastic difference between $f^{(T)}$ and~$f^{(F)}$ is that realisations drawn according to the former come
from the range of the projection~$\pi_Y$, but realisations of the latter come from the domain of~$\pi_Y$.
That is, the realisations come from different sides of equation~(\ref{Eq:Projection}).
In particular, an update to a realisation $(\mathbf{y}_{obs}, \mathbf{y}_{mis}, \mathbf{r})$ according to~$f^{(F)}$
has the form of a three tuple~$(\mathbf{y}_{obs}, \mathbf{y}^*_{mis}, \mathbf{r})$ with the {missingness}
pattern~$\mathbf{r}$ remaining unchanged.
However, an update to the same realisation $(\mathbf{y}_{obs}, \mathbf{y}_{mis}, \mathbf{r})$ according
to~$f^{(T)}$ has the form of a two tuple~$(\mathbf{y}_{obs}, \mathbf{y}^*_{mis})$, and to maintain consistency
with~$(Y, R)$, a subsequent updating of the {missingness} pattern $\mathbf{r}$ to $\mathbf{r}^*$ according to the
{missingness} mechanism~$g(\mathbf{r} |\,\mathbf{y})$ is required to complete the
triple~$(\mathbf{y}_{obs}, \mathbf{y}^*_{mis}, \mathbf{r}^*)$.

Due to this stochastic difference between $f^{(T)}$ and~$f^{(F)}$, it is important to emphasise that the correct
statement of equation~(\ref{Eq:WrongMARIdentity}) is that:
\begin{equation}
   p(\mathbf{y}_{mis} |\, \mathbf{y}_{obs}, \mathbf{r}) \,=\,
      f^{(F)}(\mathbf{y}_{mis} |\, \mathbf{y}_{obs}).
   \label{Eq:RightMARIdentity}
\end{equation}

% ==========================================================================
\section{Conceptual difficulties for the reader} \label{Sect:Difficulties}
% ==========================================================================
The difference between $f^{(T)}$ and $f^{(F)}$ and the failure in the literature to distinguish between these densities
and between variables which are formally and temporally missing creates unnecessary potential conceptual difficulties
for a reader, and this can make it difficult for a reader to obtain a coherent conceptual picture of how the related
statistical methods work. We outline some of these difficulties below.

\vspace*{2mm}
\noindent
\textbf{Difficulty~1.}
The construction of the distribution $f(\mathbf{y}_{mis} |\, \mathbf{y}_{obs})$ requires identification of variables
$\mathbf{y}_{obs}$ and $\mathbf{y}_{mis}$ in the domain of the marginal density~$f(\mathbf{y})$ for~$Y$, and
this requires the reader to deal with two inconsistent definitions of missingness simultaneously that are not distinguished
in the notation: \textit{temporal}, $(\mathbf{y}_{obs}, \mathbf{y}_{mis})$, pertaining to the marginal distribution
for~$Y$ and \textit{formal}, $(\mathbf{y}_{obs}, \mathbf{y}_{mis}, \mathbf{r})$, as defined by~$(Y, R)$.
\hfill $\qedsymbol$

\vspace*{2mm}
Note that difficulty~1 is not due to the encoding of `observed' and `missing' into the  labels `$obs$' and~`$mis$', as
opposed to~`$(1)$' and~`$(0)$' used in Rubin\;(1976), but rather because the same labels are used both in the
domain and in the range of the projection~$\pi_Y$ in equation~(\ref{Eq:Projection}).

\vspace*{2mm}
\noindent
\textbf{Difficulty~2.}
If $(\widetilde{\mathbf{y}}_{obs}, \widetilde{\mathbf{y}}_{mis}, \widetilde{\mathbf{r}})$ denotes the particular
realised values of~$(Y, R)$, then the distribution $f^{(T)}(\mathbf{y}_{mis} |\, \widetilde{\mathbf{y}}_{obs})$ is the
wrong distribution conceptually for ignorable multiple imputation.
\hfill $\qedsymbol$

\vspace*{2mm}
As we noted in Section~\ref{Sect:TwoDifferentFunctions}, an update to $\mathbf{y}_{mis}$ according
to~$f^{(T)}(\mathbf{y}_{mis} |\, \widetilde{\mathbf{y}}_{obs})$ arises as a two
tuple~$(\mathbf{y}_{obs}, \mathbf{y}^*_{mis})$ and requires an update to the {missingness} pattern to form a
completed three tuple~$(\mathbf{y}_{obs}, \mathbf{y}^*_{mis}, \mathbf{r}^*)$ to maintain consistency
with~$(Y, R)$.
Therefore, a sequence of imputations drawn according
to~$f^{(T)}(\mathbf{y}_{mis} |\, \widetilde{\mathbf{y}}_{obs})$ that is consistent with~$(Y, R)$ has the form:
\begin{equation}
   (\widetilde{\mathbf{y}}_{?},\, \mathbf{y}^{(1)}_{?},\, \mathbf{r}^{(1)}), \;
      (\widetilde{\mathbf{y}}_{?},\, \mathbf{y}^{(2)}_{?},\, \mathbf{r}^{(2)}), \;
      \ldots,
      (\widetilde{\mathbf{y}}_{?},\, \mathbf{y}^{(m)}_{?},\, \mathbf{r}^{(m)}).
   \label{Eq:BadImputations}
\end{equation}
This contrasts with a sequence of imputations drawn according
to~$f^{(F)}(\mathbf{y}_{mis} |\, \widetilde{\mathbf{y}}_{obs})$ which conceptually has the correct form:
\begin{equation}
   (\widetilde{\mathbf{y}}_{obs},\, \mathbf{y}^{(1)}_{mis},\, \widetilde{\mathbf{r}}), \;
      (\widetilde{\mathbf{y}}_{obs},\, \mathbf{y}^{(2)}_{mis},\, \widetilde{\mathbf{r}}), \;
      \ldots,
      (\widetilde{\mathbf{y}}_{obs},\, \mathbf{y}^{(m)}_{mis},\, \widetilde{\mathbf{r}}).
   \label{Eq:GoodImputations}
\end{equation}

\vspace*{2mm}
\noindent
\textbf{Difficulty~3.}
Standard conventions for interpreting mathematical notation leads to `$f$'  in~the notation
`$f(\mathbf{y}_{mis} |\, \widetilde{\mathbf{y}}_{obs})$' being interpreted as the density $f^{(T)}$ and
\textbf{not} the density $f^{(F)}$ as is required by equation~(\ref{Eq:WrongMARIdentity}) (see
equation~(\ref{Eq:RightMARIdentity})).
\hfill $\qedsymbol$

\vspace*{2mm}
Note that difficulty~3 does not apply to equation~(\ref{Eq:WrongMARIdentity}) because the context allows the
reader to interpret the function on the right hand side correctly as $f^{(F)}$ (if the reader examines the notation
carefully).
However, this is definitely \textbf{not} the case with the standalone
notation~`$f(\mathbf{y}_{mis} |\, \mathbf{y}_{obs})$',
and it is this latter notation which permeates much of the published literature on ignorable multiple imputation
methodology.

\vspace*{2mm}
\noindent
\textbf{Difficulty~4.}
Failure to distinguish between formal and temporal missingness clashes with the standard statistical convention of
inferring the identity of a density function through the denotation of the variables in its domain.
\hfill $\qedsymbol$

\vspace*{2mm}
It is common to infer from the notation `$f(x_1, x_2)$' for a joint density that `$f(x_2)$' denotes a marginal density.
However, the notation `$p(\mathbf{y}_{obs}, \mathbf{r})$' is ambiguous because the interpretation of
$\mathbf{y}_{obs}$ as formally observed leads to one function~$p$, but the interpretation of $\mathbf{y}_{obs}$
as temporally observed leads to a different function~$p$ with a different domain.

% ==========================================================================
\section{Additional limitations and notational inconsistencies} \label{Sect:AdditionalLimitations}
% ==========================================================================
Omitting from the notation the dependence of `$obs$' and `$mis$' on a specific {missingness} pattern~$\mathbf{r}$
implicitly assumes that~$\mathbf{r}$ is the only {missingness} pattern of interest to the reader.
This prevents the expression of the mathematical relationships between {missingness} patterns that exist within
equation~(\ref{Eq:WrongMARIdentity}).
Understanding these relationships at a conceptual level is useful for a reader to comprehend the primary implications
of a MAR assumption in practice where one {missingness} pattern per unit is observed, and several different
{missingness} patterns are realised overall.

The use of uppercase letters to denote both variable realisations of random vectors as well as the random vectors
themselves is common in the literature on incomplete data methods. This is contrary to the recommendations in
Halperin\;et.\,al.\;(1965). It is also another potential source of conceptual confusion for readers because the notation
`$f(Y)$' ordinarily would be understood to mean the composition~$f\circ Y$ of the density function $f$ with the
random variable~$Y$, whereas a densitiy function is something that is integrated to calculate probabilities for~$Y$.

The use of a capital `$P$' to denote a probability density function also seems fairly common in the literature on
methods for incomplete data. This too is contrary to widely understood usage of the notation where a capital $P$
denotes the probability measure and is a function of events (subsets of outcomes), whereas the density is a
corresponding function of outcomes which is integrated over subsets to calculate values for~$P$. This is a further
potential source of confusion for readers.

% ==========================================================================
\section{A remedy} \label{Sect:Remedy}
% ==========================================================================
Missing data is a common problem across a broad range of medical and public health research, and in other fields of
empirical research as well. Consequently, there is a broad range of stakeholders with an interest in being able to read
and understand the literature on the relevant statistical methods. The defects and ambiguities in the notation which we
have identified potentially undermines its purpose to disseminate the requisite information in a clear and logically
coherent manner to a {broad} range of stakeholders, because these limitations likely make this literature difficult, if not
impossible, for certain subsets of these stakeholders to read.  

Fortunately the remedy is straightforward. What is needed are four symbols rather than two to denote the partition
of $Y$ into observable and unobservable components; one pair each for the two different relationships between $Y$
and~$R$. Additionally, the dependence of the partition on a definite {missingness} pattern needs to be made explicit.
Notation for this purpose is defined formally in Appendix\;A. In practice, all one needs to understand is that four
symbols are needed
\begin{alignat*}{2}
   &Y^{ob(\mathbf{r})} \text{ and } Y^{mi(\mathbf{r})}\qquad &&(\text{to denote formal missingness}), \\
   &Y^{ot(\mathbf{r})} \text{ and } Y^{mt(\mathbf{r})}\qquad &&(\text{to denote temporal missingness}).
\end{alignat*}
Demonstrations of how this allows the difficulties discussed in Section~\ref{Sect:Difficulties} and the limitations
in Section~\ref{Sect:AdditionalLimitations} to be overcome are given in Appendices B to\;E.

% ==========================================================================
\section{Some further remarks} \label{Sect:Remarks}
% ==========================================================================
As in Seaman\;et.\,al.\,(2013), we have retained Rubin's (1976) original notation~`$f$' and~`$g$' for the marginal
density and {missingness} mechanism in a selection model factorization $f(\mathbf{y})\,g(\mathbf{r} |\, \mathbf{y})$ of
the full density. We have also retained the lowercase `$p$' from Molenberghs\;et.\,al.\,(2015\;p\,95) for the factors of
the pattern-mixture factorization~$p(\mathbf{r})\,p(\mathbf{y} |\, \mathbf{r})$, but we have introduced~`$h$' for
the full densities involving both $Y$ and~$R$, because we feel that this is clearer than denoting every density with
the generic symbol~`$p$'.

Our notation resolves the ambiguity present in `$f(\mathbf{y}_{mis}|\,\mathbf{y}_{obs})$' because by the
defintions of~$\mathbf{y}^{ob(\mathbf{r})}$ and~$\mathbf{y}^{mi(\mathbf{r})}$, the notation
`$f(\mathbf{y}^{mi(\mathbf{r})}|\,\mathbf{y}^{ob(\mathbf{r})})$' can denote only the function~$f^{(F)}$
in~(\ref{Eq:Rightf}) and \textbf{not} the function~$f^{(T)}$ in~(\ref{Eq:Wrongf}).
Alternatively, through use of the notation`$f(\mathbf{y}^{mt(\mathbf{r})} |\, \mathbf{y}^{ot(\mathbf{r})})$' 
the marginal distribution for $Y$ is freed from $(Y, R)$ in a way that does not conflict with the stochastic relationship
imposed by~$(Y, R)$.
Note that we would \text{not} expect a casual reader of the literature to understand the distinction between these
notations, and we recommend that authors \textbf{always} explain that imputations drawn according to
$f(\mathbf{y}^{mi(\mathbf{r})}|\,\mathbf{y}^{ob(\mathbf{r})})$ represent triples
$(\widetilde{\mathbf{y}}^{ob(\mathbf{r})}, \mathbf{y}^{mi(\mathbf{r})}, \widetilde{\mathbf{r}})$ from
$\Omega_{\widetilde{\mathbf{r}}}$ and \textbf{not} pairs
$(\widetilde{\mathbf{y}}^{ot(\mathbf{r})}, \mathbf{y}^{mt(\mathbf{r})})$ from~$\mathcal{Y}$.

The additional ambiguities noted earlier are resolved in our notation as well.
For example, the two functions corresponding to the notation `$p(\mathbf{y}_{obs}, \mathbf{r})$' are
denoted by $h(\mathbf{y}^{ob(\mathbf{r})}, \mathbf{r})$
and~$h(\mathbf{y}^{ot(\mathbf{r}_j)}, \mathbf{r})$, respectively, in our notation
(see equations~(\ref{Eq:ObservableDataComponentPM1}) and~(\ref{Eq:ObservableDataComponentPM2}) in
Appendix~B). We have not addressed explicity the notation `$p(\mathbf{r} |\, \mathbf{y}_{obs})$' because this is
the subject of a separate investigation {currently in preparation(delete) treated in detail in Galati\;(2019)}.

We have maintained consistency with the recommendation in Halperin\;et.\,al.\,(1965) to distinguish between random
vectors and their realisations with uppercase and lowercase letters, respectively. However, we saw no need to consider
vectors specifically to be column matrices in the present circumstances. Also, one non-standard feature of our notation
is that we use $\Omega$ to denote the range of the random vector $(Y, R)$ instead of its domain. This is because we
have a need explicitly to refer to subsets of~$\Omega$, and in likelihood theory the probability spaces of interest
typically are defined entirely on the range of $(Y, R)$ via density functions (and often there is no need to refer
explicitly to some underlying sample space).
 
We hope that identification and elaboration of the notational issues raised in this paper will assist readers to navigate
more easily the existing literature on statistical methods for incomplete data, and to assist future authors to improve the
clarity of their expositions.

% ==========================================================================
\section{Appendix A (Definition of Notation)} \label{Sect:Notation}
% ==========================================================================

% ----------------------------------------------------------------------------------------------
\subsection{Random Vectors}
    \label{Sect:RandomVectors}
% ----------------------------------------------------------------------------------------------

Throughout, $Y$ denotes a random vector modelling the observed and unobserved data comprising all units in the
study jointly, and $R$ denotes a random vector of binary response random variables of the same dimension as~$Y$,
where `1' means observed.
Joint distributions for the pair of random vectors $(Y, R)$ will be referred to as \textbf{full} distributions.

\vspace*{2mm}
\noindent
\textbf{Note.}
We have no need to distinguish between vectors interpreted as column matrices versus row matrices, and so for our
purposes we do not give vectors column matrix interpretation and dispense with the common~`$'$' and~`$^T$'
notations.

\vspace*{2mm}
\noindent
\textbf{Note.}
Typically a data analyst thinks of a given $\mathbf{y}$ as comprising a rectangular matrix with each column
pertaining to a specific `variable' (for example, blood pressure) and each row pertaining to a specific unit
(for example, an individual in the study).
In our notation,
the data matrix is shaped so that there is a single row with the data for the various units placed side by side in
sets of columns.

% ----------------------------------------------------------------------------------------------
\subsection{Sample Spaces}
    \label{Sect:SampleSpaces}
% ----------------------------------------------------------------------------------------------
Let $\mathcal{R} = \{\mathbf{r}_1, \mathbf{r}_2, \ldots, \mathbf{r}_k \}$ be the set of distinct {missingness} patterns
with $\mathbf{r}_1 = \mathbf{1}$ denoting the `all ones' vector corresponding to the complete cases.
For convenience, we let $\mathbf{r}_0 = \mathbf{0}$ denote the `all zeros' vector corresponding to
non-participants, where it may or may not be the case that $\mathbf{r}_j = \mathbf{r}_0$ for some
$j\in\{1, 2, \ldots, k\}$.
(We exclude $j=0$ so as to avoid ever having $P(\mathbf{r}_0) = 0$.)
Note that the dot product $\mathbf{r}_j\boldsymbol{\cdot}\mathbf{r}_j$ gives the number of values observed
when the $j^{th}$ {missingness} pattern is realised and, in particular, $\mathbf{r}_1\boldsymbol{\cdot}\mathbf{r}_1$
gives the number of variables in~$R$ (and also in~$Y$).
Let $\mathcal{Y} = \text{range}(Y)$ be the set of realisable datasets, where a \textbf{realizable} dataset contains
complete data including all values that may or may not be observable.

Let
$\Omega = \mathcal{Y}\times\mathcal{R} =
   \Omega_1 \;\dot{\cup}\; \Omega_2 \;\dot{\cup}\; \cdots \;\dot{\cup}\; \Omega_k$
be the \textbf{full sample space} of realisable pairs of datasets and {missingness} patterns,
where $\Omega_j = \mathcal{Y}\times\{\mathbf{r}_j\}$ for~$\mathbf{r}_j\in\mathcal{R}$.
When the subscript $j$ of $\mathbf{r}$ is omitted, we denote $\Omega_j$ by~$\Omega_\mathbf{r}$.
Let $\pi_Y$ and $\pi_R$ denote the projections $(\mathbf{y}, \mathbf{r})\mapsto\mathbf{y}$ and
$(\mathbf{y}, \mathbf{r})\mapsto\mathbf{r}$, respectively.

Realisations which represent a specific realisable dataset or {missingness} pattern only are denoted
$\widetilde{\mathbf{y}}$ and~$\widetilde{\mathbf{r}}$, respectively.

% ----------------------------------------------------------------------------------------------
\subsection{Projections on $\mathcal{Y}$ and~$\Omega_j$}
    \label{Sect:Projections}
% ----------------------------------------------------------------------------------------------
For $j=1, 2, \ldots, k$, let $\pi(\mathbf{r}_j)\,:\,\mathcal{Y}\rightarrow\mathcal{Y}^{\pi(\mathbf{r}_j)}$ and
$\pi(\neg\mathbf{r}_j)\,:\,\mathcal{Y}\rightarrow\mathcal{Y}^{\pi(\neg\mathbf{r}_j)}$
denote the projections extracting from each $\mathbf{y}$ vector the vectors of its observed and unobserved
values, respectively, according to the {missingness} pattern $\mathbf{r}_j$.
(In logic, `$\neg$' is commonly used for negation.)
By convention we set $\pi(\mathbf{r}_0) = \pi(\neg\mathbf{r}_1) = \varnothing$.
To apply these projections correctly over $\Omega$, we define the following mappings
\begin{alignat}{1}
   o\,:\,\mathcal{R} &\rightarrow
      \{ \pi(\mathbf{r})\circ\pi_Y : \Omega_{\mathbf{r}}\rightarrow\mathcal{Y}^{\pi(\mathbf{r})}\}
   \label{Eq:ObMap} \\
   m\,:\,\mathcal{R} &\rightarrow
      \{ \pi(\neg\mathbf{r})\circ\pi_Y :
      \Omega_{\mathbf{r}}\rightarrow\mathcal{Y}^{\pi(\neg\mathbf{r})} \}
   \label{Eq:MiMap}
\end{alignat}
and use an abbreviated notation to refer to the images of $(\mathbf{y}, \mathbf{r})\in\Omega$ under these
mappings:
\begin{alignat}{1}
   \mathbf{y}^{ob(\mathbf{r})} \,&:=\,
      (\mathbf{y}, \mathbf{r})^{o(\pi_R(\mathbf{y}, \mathbf{r}))}
   \label{Eq:YObR} \\
   \mathbf{y}^{mi(\mathbf{r})} \,&:=\,
      (\mathbf{y}, \mathbf{r})^{m(\pi_R(\mathbf{y}, \mathbf{r}))}.
   \label{Eq:YMiR}
\end{alignat}
Additionally, for $\mathbf{r}\in\mathcal{R}$ and $\mathbf{y}\in\mathcal{Y}$ set
\begin{alignat}{1}
   \mathbf{y}^{ot(\mathbf{r})} \,&:=\, 
   \begin{cases} 
      \mathbf{y}^{\pi(\mathbf{r})}
         & \text{over } \mathcal{Y} \\
      (\mathbf{y}, \mathbf{r}_j)^{o(\pi_R(\mathbf{y}, \mathbf{r}))}
         & \text{over } \mathcal{Y}\times\mathcal{R}
   \end{cases}
   \label{Eq:YObtR} \\
   \mathbf{y}^{mt(\mathbf{r})} \,&:=\, 
   \begin{cases} 
      \mathbf{y}^{\pi(\neg\mathbf{r})}
         & \text{over } \mathcal{Y} \\
      (\mathbf{y}, \mathbf{r}_j)^{m(\pi_R(\mathbf{y}, \mathbf{r}))}
         & \text{over } \mathcal{Y}\times\mathcal{R}.
   \end{cases}
   \label{Eq:YMitR}
\end{alignat}

\vspace*{2mm}
\noindent
\textbf{Note.}
The notations in (\ref{Eq:ObMap}) and (\ref{Eq:MiMap}) and on the right hand sides of
(\ref{Eq:YObR})$-$(\ref{Eq:YMitR}) may seem unwieldy.
Note that these notations are needed solely for the purpose of carefully defining the four symbols
$\mathbf{y}^{ob(\mathbf{r})}$, $\mathbf{y}^{mi(\mathbf{r})}$, $\mathbf{y}^{ot(\mathbf{r})}$
and~$\mathbf{y}^{mt(\mathbf{r})}$.
It is only these latter four symbols that are needed for working with densities for the distributions for~$(Y, R)$
themselves.

\vspace*{2mm}
\noindent
\textbf{Note.}
The vectors $\mathbf{y}^{ob(\mathbf{r})}$ and $\mathbf{y}^{ot(\mathbf{r})}$ have length
$\mathbf{r}\boldsymbol{\cdot}\mathbf{r}$ while the vectors $\mathbf{y}^{mi(\mathbf{r})}$ and
$\mathbf{y}^{mt(\mathbf{r})}$ have length
$\mathbf{r}_1\boldsymbol{\cdot}\mathbf{r}_1 - \mathbf{r}\boldsymbol{\cdot}\mathbf{r}$.
Note that these lengths vary from {missingness} pattern to {missingness} pattern.

\vspace*{2mm}
\noindent
\textbf{Note.}
The projections $ob(\mathbf{r})$ and $mi(\mathbf{r})$ apply \textit{solely} on the range of $(Y, R)$ and are
\textit{always} consistent with the missingness relation~$(Y, R)$.
Each {missingness} pattern $\mathbf{r}_j$ gives projections~$ob(\mathbf{r}_j)$ and~$mi(\mathbf{r}_j)$ on~$\Omega_j$,
and {each(delete) these} are pieced together over all {missingness} patterns to {give} a single pair of functions on all of~$\Omega$. 

\vspace*{2mm}
\noindent
\textbf{Note.}
The projections $ot(\mathbf{r})$ and $mt(\mathbf{r})$ apply on either $Y$ or $(Y, R)$ as the context dictates.
Each $\mathbf{r}_j$ gives a distinct pair of projections $ot(\mathbf{r}_j)$ and $mt(\mathbf{r}_j)$ on \textit{all}
of~$\mathcal{Y}$ or \textit{all} of~$\Omega$, as the case may be.
In the latter case, {these(delete)} $ot(\mathbf{r}_j)$ and $mt(\mathbf{r}_j)$ are consistent with~$(Y, R)$ on $\Omega_j$
and \textbf{inconsistent} with~$(Y, R)$ elsewhere on~$\Omega$.
The `$t$' in `$ot$' and `$mt$' can be taken to mean `temporally' or `this time'.

\vspace*{2mm}
\noindent
\textbf{Note.}
The notation `$f(\mathbf{y}^{mt(\mathbf{r})} |\, (\mathbf{y}^{ot(\mathbf{r})})$' is ambiguous because as
defined by~(\ref{Eq:YObtR}) and~(\ref{Eq:YMitR}) this can denote either the function~$f^{(T)}$
(see~(\ref{Eq:Wrongf})) defined on~$\mathcal{Y}$ or a function (not~$f^{(F)}$) defined on all of~$\Omega$
{, but which one is intended should be clear from the context}.
However, the notation `$f(\mathbf{y}^{mi(\mathbf{r})} |\, (\mathbf{y}^{ob(\mathbf{r})})$' is \textbf{unambiguous}
because by~(\ref{Eq:YObR}) and~(\ref{Eq:YMiR}) this must denote~$f^{(F)}$.

% ----------------------------------------------------------------------------------------------
\subsection{Observable Data Events}
    \label{Sect:ODEs}
% ----------------------------------------------------------------------------------------------
Given $(\mathbf{y}, \mathbf{r})\in\Omega$, we call
\begin{equation}
   \Omega_{(\mathbf{y},\mathbf{r})} =
      \{\, (\mathbf{y}_*, \mathbf{r})
      \,:\, \mathbf{y}_*^{ob(\mathbf{r})}=\mathbf{y}^{ob(\mathbf{r})} \,\} \subset \mathcal{Y}\times\mathcal{R}
   \label{Eq:ODE}
\end{equation}
the \textbf{observed data event} corresponding to~$(\mathbf{y}, \mathbf{r})$.
The set $\Omega_{(\mathbf{y},\mathbf{r})}$ consists of all datasets
$\mathbf{y}_*$ which have the same observed values as~$\mathbf{y}$ (as defined by the {missingness}
pattern~$\mathbf{r}$).
For a fixed~$\mathbf{r}\in\mathcal{R}$, the events in~(\ref{Eq:ODE}) partition~$\Omega_\mathbf{r}$, and over
all~$\mathbf{r}$ they give a partition of~$\Omega$.
These \textbf{observable data events} are the classes of the equivalence relation defined by setting for all
$(\mathbf{y}_1, \mathbf{r}_1), (\mathbf{y}_2, \mathbf{r}_2) \in \Omega$,
$(\mathbf{y}_1, \mathbf{r}_1) \sim_\text{ob} (\mathbf{y}_2, \mathbf{r}_2)$ if, and only if,
$\mathbf{r}_1 = \mathbf{r}_2$ and
$\mathbf{y}_1^{ob(\mathbf{r}_1)} = \mathbf{y}_2^{ob(\mathbf{r}_2)}$.

% ----------------------------------------------------------------------------------------------
\subsection{Density Functions}
    \label{Sect:Densities}
% ----------------------------------------------------------------------------------------------
We specify full distributions for $(Y, R)$ through density functions $h:\Omega\rightarrow\mathbb{R}$,
with probabilities determined by integration: $P(A) = \int_A h$ for any $A\subseteq\mathcal{Y}\times\mathcal{R}$
for which a probability can be defined (see Ash and Dol\'{e}ans-Dade\;(2000) or~Shorack\;(2000) for details). Note
{that} we suppress the dominating measure in the notation. Two different ways of factorizing $h$ are useful:
\begin{equation}
   h(\mathbf{y}, \mathbf{r}) \; = \;
      f(\mathbf{y})\,g(\mathbf{r}\,|\,\mathbf{y}) \; = \;
      p(\mathbf{r})\,p(\mathbf{y}\,|\,\mathbf{r})
   \label{Eq:FullDensity}
\end{equation}
for all $(\mathbf{y},\mathbf{r})\in\mathcal{Y}\times\mathcal{R}$. The first factorization is called a \textbf{selection
model} factorization of~$h$, and the factor $g(\mathbf{r}\,|\,\mathbf{y})$ is called a \textbf{missingness mechanism}.
The second factorization is called a \textbf{pattern-mixture} factorization, and for each~$\mathbf{r}\in\mathcal{R}$,
we call the conditional density $p(\mathbf{y}\,|\,\mathbf{r})$ the \textbf{pattern mixture component} pertaining
to~$\mathbf{r}$.

\vspace*{2mm}
\noindent
\textbf{Note.}
As specified in~(\ref{Eq:FullDensity}), a missingness mechanism $g(\mathbf{r}\,|\,\mathbf{y})$ is a function of two
vector variables $\mathbf{y}$ and $\mathbf{r}$ defined on all of $\mathcal{Y}\times\mathcal{R}$ subject to the
restrictions that
$0\leq g(\mathbf{r}\,|\,\mathbf{y})\leq 1$ for each $(\mathbf{y},\mathbf{r})\in\mathcal{Y}\times\mathcal{R}$
and $\sum_{i=1}^{k}g(\mathbf{r}_i\,|\,\mathbf{y})=1$ for each fixed $\mathbf{y}\in\mathcal{Y}$.
We stress that instead of the ususal interpretation of considering a missingness mechanism to give a conditional
probability distribution for~$R$ for each fixed $\mathbf{y}\in\mathcal{Y}$, the perspective that will be relevant for
us is to consider the behaviour of $g$ as a mathematical function of both $\mathbf{y}$ and $\mathbf{r}$ when its
domain, $\mathcal{Y}\times\mathcal{R}$, is restricted to an observed data event
$\Omega_{(\mathbf{y},\mathbf{r})}\subset\mathcal{Y}\times\mathcal{R}$,
that is, to a set of the form~(\ref{Eq:ODE}). This perspective is specific to the incomplete data setting, and typically
does not arise with complete-data statistical methods.
\hfill $\qedsymbol$

\vspace*{2mm}
\noindent
\textbf{Note.}
Technically, the symbols $h$, $f$, $g$ and $p$ denote functions and $h(\mathbf{y}, \mathbf{r})$,
$f(\mathbf{y})$, $g(\mathbf{r}\,|\,\mathbf{y})$, $p(\mathbf{r})$ and $p(\mathbf{y}\,|\,\mathbf{r})$ denote real
numbers.
Because it is common in statistics to use the same symbol to denote different densities, for example a joint
density $f(x_1, x_2)$ and a marginal density~$f(x_1)$, we adopt the usual convention and often refer to these
functions by their values.
\hfill $\qedsymbol$

% ==========================================================================
\section{Appendix B (The observable data distribution)} \label{Sect:ObservableDataDistribution}
% ==========================================================================
To apply likelihood theory to incomplete data, from the model for the full data one must construct a model for just the
observable data. This involves specifying a set of outomes and a set of events for the observable data, and to each full
density~$h$, a corresponding density on the set of outcomes for the observable data. Here we give an explicit 
construction for this probability space together with a step-by-step derivation of the density given in~(5.11) in the
extract quoted in Section~\ref{Sect:Introduction}.

The outcomes can be taken to be either the set of observable data events or the range of the map
$(\textbf{y}, \textbf{r})\mapsto(\mathbf{y}^{ob(\mathbf{r})}, \mathbf{r})$ because there is a one-to-one
correspondence between~$\{\Omega_{(\mathbf{y},\mathbf{r})}\}$
and~$\{(\mathbf{y}^{ob(\mathbf{r})}, \mathbf{r})\}$.
The latter seems to be preferred~(Little and Rubin\;(1987), Little and Rubin\;(2002), Tsiatis\;(2006)):
\begin{equation}
   \Omega_{ob} \,:=\,
      \bigcup_{j=1}^k \left(\mathcal{Y}^{ot(\mathbf{r}_j)}\times\{\,\mathbf{r}_j\,\}\right).
   \label{Eq:ObsSampleSpace}
\end{equation}
This is an irregularly-shaped set because as noted in Appendix~A the vectors
$\mathbf{y}^{ot(\mathbf{r}_j)}$ typically have different lengths for different {missingness} patterns.

Under the one-to-one correspondence between~$\{(\mathbf{y}^{ob(\mathbf{r})}, \mathbf{r})\}$
and~$\{\Omega_{(\mathbf{y},\mathbf{r})}\}$, events in $\Omega_{ob}$
correspond to unions of observable data events in~$\Omega$.
Restricting to observable data events gives the density for the probability distribution on~$\Omega_{ob}$:
\begin{equation}
   h(\mathbf{y}^{ob(\mathbf{r})}, \mathbf{r}) \,=\,
      \int f(\mathbf{y})\, g(\mathbf{r} |\, \mathbf{y})\,
         \text{d}\mathbf{y}^{mi(\mathbf{r})} \,=\,
      \int_{\Omega_{(\mathbf{y},\mathbf{r})}}
         \mspace{-25mu}h(\mathbf{y}, \mathbf{r}).
   \label{Eq:ObservableDataComponentSM}
\end{equation}
This can be seen to be the required density simply by pulling events in $\Omega_{ob}$ back to unions of observable 
data events in~$\Omega$ and integrating $h$ over these corresponding events for~$(Y, R)$ by applying iterated
integrals as per Fubini's Theorem (Ash and Dol\'{e}ans-Dade (2000\;p\,101)). Note that we use $mi(\mathbf{r})$ and
not $mt(\mathbf{r})$ in `$\text{d}\mathbf{y}^{mi(\mathbf{r})}$' because the integrand $h$  is defined on all of
$\Omega$ and the variables integrated out of $h$ are different for each
subset~$\Omega_j$.

The {missingness}-pattern-dependant processing being performed in the construction of the density
in~(\ref{Eq:ObservableDataComponentSM}) does not correlate well with the selection-model factorization for~$h$,
and this can make the construction seem a little opaque. An alternative derivation is possible starting with a
pattern-mixture factorization for~$h$.

One way to do this is to start from
$h(\mathbf{y}, \mathbf{r}) =
   p(\mathbf{r}) \, p(\mathbf{y}^{mi(\mathbf{r})}, \mathbf{y}^{ob(\mathbf{r})} |\, \mathbf{r})$,
restrict $\Omega$ to~$\Omega_j$:
$h(\mathbf{y}, \mathbf{r}_j) =
      p(\mathbf{r}_j) p(\mathbf{y}^{mi(\mathbf{r}_j)}, \mathbf{y}^{ob(\mathbf{r}_j)} |\, \mathbf{r}_j)$,
marginalize to~$\mathbf{y}^{ob(\mathbf{r}_j)}$:
\begin{alignat}{1}
   h(\mathbf{y}^{ob(\mathbf{r}_j)}, \mathbf{r}_j) \,&=\,
      \int
      p(\mathbf{r}_j) \,
      p(\mathbf{y}^{mi(\mathbf{r}_j)}, \mathbf{y}^{ob(\mathbf{r}_j)} |\, \mathbf{r}_j)  \,
      \text{d}\mathbf{y}^{mi(\mathbf{r}_j)}
      \notag \\
   \,&=\,
      p(\mathbf{r}_j)
     \int
      p(\mathbf{y}^{mi(\mathbf{r}_j)}, \mathbf{y}^{ob(\mathbf{r}_j)} |\, \mathbf{r}_j)  \,
      \text{d}\mathbf{y}^{mi(\mathbf{r}_j)}
      \notag \\
   \,&=\,
      p(\mathbf{r}_j) \,
      p(\mathbf{y}^{ob(\mathbf{r}_j)} |\, \mathbf{r}_j)
      \label{Eq:ObservableDataComponentPM1}
\end{alignat}
and then put the pieces together over all of~$\Omega_{ob}$:
$h(\mathbf{y}^{ob(\mathbf{r})}, \mathbf{r}) =
   p(\mathbf{r}) \,   p(\mathbf{y}^{ob(\mathbf{r})} |\, \mathbf{r})$.
Alternatively, for each $j$ one can marginalize over all of~$\Omega$:
\begin{alignat}{1}
   h(\mathbf{y}^{ot(\mathbf{r}_j)}, \mathbf{r}) \,&=\,
      \int
      p(\mathbf{r}) \,
      p(\mathbf{y}^{mt(\mathbf{r}_j)}, \mathbf{y}^{ot(\mathbf{r}_j)} |\, \mathbf{r}) \,
      \text{d}\mathbf{y}^{mt(\mathbf{r}_j)}
      \notag \\
   \,&=\,
      p(\mathbf{r})
      \int
      p(\mathbf{y}^{mt(\mathbf{r}_j)}, \mathbf{y}^{ot(\mathbf{r}_j)} |\, \mathbf{r}) \,
      \text{d}\mathbf{y}^{mt(\mathbf{r}_j)}
      \notag \\
   \,&=\,
      p(\mathbf{r}) \,
      p(\mathbf{y}^{ot(\mathbf{r}_j)} |\, \mathbf{r}),
      \label{Eq:ObservableDataComponentPM2}
\end{alignat}
restrict to~$\Omega_j$:
$h(\mathbf{y}^{ot(\mathbf{r}_j)}, \mathbf{r}_j) =
   h(\mathbf{y}^{ob(\mathbf{r}_j)}, \mathbf{r}_j) =
   p(\mathbf{r}_j) p(\mathbf{y}^{ob(\mathbf{r}_j)} |\, \mathbf{r}_j)$,
and then put the pieces together over all of~$\Omega_{ob}$:
\begin{equation}
   h(\mathbf{y}^{ob(\mathbf{r})}, \mathbf{r}) \,=\,
      p(\mathbf{r}) \,
      p(\mathbf{y}^{ob(\mathbf{r})} |\, \mathbf{r}).
      \label{Eq:ObservableDataDensitytPM}
\end{equation}

\vspace*{2mm}
\noindent
\textbf{Note.}
In~(\ref{Eq:ObservableDataComponentPM2}), for a given $j$ the density
$h(\mathbf{y}^{ot(\mathbf{r}_j)}, \mathbf{r})$ is a marginal density of~$h$ with
domain~$\mathcal{Y}^{ot(\mathbf{r}_j)}\times\mathcal{R}$. 
There are $k$ of these distributions.
On the other hand, there is only one density $h(\mathbf{y}^{ob(\mathbf{r})}, \mathbf{r})$ with
domain~$\Omega_{ob}$.
For a given~$j$, the function $h(\mathbf{y}^{ot(\mathbf{r}_j)}, \mathbf{r})$ agrees with
$h(\mathbf{y}^{ob(\mathbf{r})}, \mathbf{r})$ on the set~$\Omega_j$, but comparison of these two functions on the rest
of their domains is not well defined.

\vspace*{2mm}
\noindent
\textbf{Note.}
Because (\ref{Eq:ObsSampleSpace}) is irregularly shaped and not a Cartesian product, the stochastic function
obtained by composing $(Y, R)$ with $(\mathbf{y}, \mathbf{r})\mapsto(\mathbf{y}^{ob(\mathbf{r})}, \mathbf{r})$
is \textbf{not} a random vector.
Tsiatis (\cite{Tsiatis06}~page~13) calls these `random quantities'.
Stochastic functions more general than random vectors are called `random objects' by Ash and
Dol\'{e}ans-Dade~(2000\;p\,178)  and `random elements' by Shorack (2000\;p\,90). To be applicable to incomplete
data, the likelihood theory must be sufficiently general to cover these random quantities. See Shorack (2000\;pp.\,563--567)
for a sufficiently general likelihood theory for the case of IID data.

\vspace*{2mm}
\noindent
\textbf{Note.}
If $Y_{obs}$ is interpreted as formally observed and considered to vary over {missingness} patterns, then it denotes
the composition of $(Y, R)$ with $(\mathbf{y}, \mathbf{r})\mapsto\mathbf{y}^{ob(\mathbf{r})}$.
As was noted in Section~\ref{Sect:TwoNotionsOfMissingness}, when interpreted this way $Y_{obs}$ alone is
insufficent to model the observable data.
This is because there is potential for clashes between the ranges from distinct {missingness} pattern{s}.
That is, we may have $\mathbf{r}_j\neq\mathbf{r}_{j'}$ with
$\mathcal{Y}^{ot(\mathbf{r}_j)} = \mathcal{Y}^{ot(\mathbf{r}_{j'})}$ on the right hand side
of~(\ref{Eq:ObsSampleSpace}).

% ==========================================================================
\section{Appendix C (Distributions of temporally missing variables)} \label{Sect:Demonstration}
% ==========================================================================
We give a formal demonstration that the random vectors $Y^{ot(\mathbf{r})}$ and $Y^{mt(\mathbf{r})}$ each
comprise mixtures of formally observable and formlly unobservable data.
To do this neatly, we define a partial-order on~$\mathcal{R}$ as follows (see Davey and Priestly\;(2002) for the
definition of a partial order): for each $l \in \{\, 1, 2, \ldots, \mathbf{r}_1 \boldsymbol{\cdot} \mathbf{r}_1 \,\}$ let
$\pi_l$ denote the projection with domain $\mathcal{R}$ extracting the $l^{th}$ coordinate of each {missingness} pattern.
Then for $\mathbf{r}_i, \mathbf{r}_j\in\mathcal{R}$ define
\begin{equation}
   \mathbf{r}_i \le_\text{p} \mathbf{r}_j 
   \quad\Leftrightarrow\quad
   \pi_l(\mathbf{r}_i) \leq \pi_l(\mathbf{r}_j)\text{ for all } l \in
      \{\, 1, 2, \ldots, \mathbf{r}_1 \boldsymbol{\cdot} \mathbf{r}_1 \,\}.
   \label{Eq:PartialOrder}
\end{equation}
In words,
$\mathbf{r}_i \le_\text{p} \mathbf{r}_j$
if, and only if, all values that are defined to be observed according to pattern~$\mathbf{r}_i $ are defined to be
observed according to pattern~$\mathbf{r}_j$.
It is straightforward to check that this relation is reflexive, transitive and anti-symmetric.

Let $\mathbf{r}\in\mathcal{R}$ and consider a full density $h$ for $(Y, R)$ as in (\ref{Eq:FullDensity}) factored into
selection model and pattern-mixture forms,
$h(\mathbf{y}, \mathbf{r}) =
   f(\mathbf{y}) \, g(\mathbf{r}\,|\,\mathbf{y}) =
   p(\mathbf{r}) \, p(\mathbf{y}\,|\,\mathbf{r})$.
Marginalising the latter factorization over all {missingness} patterns gives the marginal density for~$Y$ as a mixture of
the pattern-mixture components:
\begin{equation}
   f(\mathbf{y}) \;\;=\;\;
      \sum_{j=1}^{k}
      p(\mathbf{r}_j)\,
      p(\mathbf{y}\,|\,\mathbf{r}_j).
      \label{Eq:MarginalYAsPatterns}
\end{equation}
Letting $\mathbf{y}\in\mathcal{Y}$ and substituting
$\mathbf{y} = (\mathbf{y}^{mt(\mathbf{r})}, \mathbf{y}^{ot(\mathbf{r})})$ into both sides
of~(\ref{Eq:MarginalYAsPatterns}) gives:
\begin{equation}
   f(\mathbf{y}^{mt(\mathbf{r})}, \mathbf{y}^{ot(\mathbf{r})}) \;\;=\;\; 
      \sum_{j=1}^{k}
      p(\mathbf{r}_j)\,
      p(\mathbf{y}^{mt(\mathbf{r})}, \mathbf{y}^{ot(\mathbf{r})} |\, \mathbf{r}_j \,).
      \label{Eq:MixedUpPatterns}
\end{equation}
Now in the sum on the right-hand side of~(\ref{Eq:MixedUpPatterns}), the terms for which all the entries of
$\mathbf{y}^{mt(\mathbf{r})}$ are labelled as formally missing according to $(Y, R)$ are those with {missingness}
patterns satisfying $\mathbf{r}_j \le_p \mathbf{r}$ (according to the partial order defined
in~(\ref{Eq:PartialOrder})).
Similarly, the terms for which all the entries of $\mathbf{y}^{ot(\mathbf{r})}$ are labelled as formally observed
according to $(Y, R)$ are those with {missingness} patterns satisfying~$\mathbf{r} \le_p \mathbf{r}_j$.
By anti-symmetry, the only component on the right-hand side of~(\ref{Eq:MixedUpPatterns}) for which all labelling
of the $\mathbf{y}$ values is formally correct is the single component with $\mathbf{r}_i = \mathbf{r}$.
Hence, provided $\mathcal{R}$ contains at least two {missingness} patterns, one of $\mathbf{y}^{ot(\mathbf{r})}$ and
$\mathbf{y}^{mt(\mathbf{r})}$ is a mixture of formally observable and formally unobservable data.
(This shows that at least one of $Y^{ot(\mathbf{r})}$ and $Y^{mt(\mathbf{r})}$ is mixed.
In most cases, this will be true of both.)

% ==========================================================================
\section{Appendix D (Derivation of the MAR Identity)} \label{Sect:Derivation}
% ==========================================================================
Here we give a formal derivation of equation~(\ref{Eq:WrongMARIdentity}). Given
$h(\mathbf{y},\mathbf{r})=f(\mathbf{y})\,g(\mathbf{r}\,|\,\mathbf{y})$ factorised in selection model form together
with observed data~$\Omega_{(\mathbf{y},\mathbf{r})}$, we say that the {missingness} mechanism
$g(\mathbf{r}\,|\,\mathbf{y})$ is \textbf{missing at random} (\textbf{MAR}) with respect to 
$\Omega_{(\mathbf{y},\mathbf{r})}$ if $g(\mathbf{r}\,|\,\mathbf{y})$ is a constant function
on~$\Omega_{(\mathbf{y},\mathbf{r})}$. Define $\mathcal{M} = \{ h_{(\theta, \psi)} \,:\, (\theta, \psi)\in\Delta\,\}$
to be MAR when MAR holds with respect to $\Omega_{(\mathbf{y},\mathbf{r})}$ for all densities $h_{(\theta, \psi)}$
in~$\mathcal{M}$. Everywhere MAR in Seaman\;et.\,al.\,(2013) is accommodated by requiring that MAR hold with
respect to all observed data events (for all densities in~$\mathcal{M}$).

\vspace*{2mm}
Let $h$ be as in~(\ref{Eq:FullDensity}) and let
$(\mathbf{y}, \mathbf{r})\in\Omega\backslash\Omega_{\mathbf{r}_1}$ be a partially-observed realisation drawn
according to~$h$.
Partitioning $\mathbf{y}$ into observable and unobservable components as defined by~$\mathbf{r}$ gives
\begin{equation}
p(\mathbf{r}) \, p(\mathbf{y}^{mi(\mathbf{r})}, \mathbf{y}^{ob(\mathbf{r})} |\, \mathbf{r}) \,=\,
   f(\mathbf{y}^{mi(\mathbf{r})}, \mathbf{y}^{ob(\mathbf{r})}) \,
   g(\mathbf{r} \,|\, \mathbf{y}).
  \label{Eq:PartitionedPMandSM}
\end{equation}

\vspace*{2mm}
\noindent
Note that the `$f$' in $f(\mathbf{y}^{ob(\mathbf{r})} |\, \mathbf{y}^{ob(\mathbf{r})})$ denotes the
function~$f^{(F)}$ (see~(\ref{Eq:Rightf})) and \textbf{not} the function~$f^{(T)}$.
Factorizing the joint density for the $\mathbf{y}$ values on each side of (\ref{Eq:PartitionedPMandSM}) into the
product of a marginal and a conditional density, and then rearranging (provided all required denominators are
non-zero) gives:
\begin{equation}
   p(\mathbf{y}^{mi(\mathbf{r})} |\, \mathbf{y}^{ob(\mathbf{r})}, \mathbf{r}) \,=\,
      \frac{
         f(\mathbf{y}^{mi(\mathbf{r})} |\, \mathbf{y}^{ob(\mathbf{r})}) \,
         f(\mathbf{y}^{ob(\mathbf{r})}) \,
         g(\mathbf{r} \,|\, \mathbf{y})}
         {p(\mathbf{r}) \, p(\mathbf{y}^{ob(\mathbf{r})} |\, \mathbf{r})}.
      \label{Eq:PMSM} 
\end{equation}
In~(\ref{Eq:PMSM}) the function $f(\mathbf{y}^{ob(\mathbf{r})})$ denotes the composition of the marginal density
$f(\mathbf{y}^{ot(\mathbf{r})})$ with the projection~$\pi_Y$ (suitably restricted).

If $g$ is MAR with respect to~$\Omega_{(\mathbf{y},\mathbf{r})}$, then
over $\Omega_{(\mathbf{y},\mathbf{r})}$ the only non-constant factor
on the right hand side is~$f(\mathbf{y}^{mi(\mathbf{r})} |\, \mathbf{y}^{ob(\mathbf{r})})$.
Integrating both sides with respect to the $\mathbf{y}^{mi(\mathbf{r})}$ variables and rearranging gives
\begin{equation}
   p(\mathbf{y}^{ob(\mathbf{r})} |\, \mathbf{r}) \,=\,
      \frac{1}{p(\mathbf{r})} \,
         f(\mathbf{y}^{ob(\mathbf{r})}) \,
         g(\mathbf{r} |\, \mathbf{y})
      \label{Eq:MARODDI} 
\end{equation}
because
$\int p(\mathbf{y}^{mi(\mathbf{r})} |\,
   \mathbf{y}^{ob(\mathbf{r})}, \mathbf{r}) \text{d}\mathbf{y}^{mi(\mathbf{r})} =
\int f(\mathbf{y}^{mi(\mathbf{r})} |\,
   \mathbf{y}^{ob(\mathbf{r})}) \text{d}\mathbf{y}^{mt(\mathbf{r})} = 1$.
Substituting (\ref{Eq:MARODDI}) back into (\ref{Eq:PMSM}) then gives 
\begin{equation}
   p(\mathbf{y}^{mi(\mathbf{r})} |\, \mathbf{y}^{ob(\mathbf{r})}, \mathbf{r}) \,=\,
      f(\mathbf{y}^{mi(\mathbf{r})} |\, \mathbf{y}^{ob(\mathbf{r})}).
      \label{Eq:RightMARIdentityII} 
\end{equation}

% ==========================================================================
\section{Appendix E (Further analysis of the MAR Identity)} \label{Sect:Analysis}
% ========================================================================== 
In this final Appendix we examine the MAR identity~(\ref{Eq:RightMARIdentityII}) more closely. For a fixed 
$(\mathbf{y}, \mathbf{r})\in\Omega$, the domain of the densities in this equality is the observed data
event~$\Omega_{(\mathbf{y},\mathbf{r})}$.
When restricted to this event, $\pi_Y$ gives a bijection onto a corresponding subset
$\pi_Y\left(\Omega_{(\mathbf{y},\mathbf{r})}\right)$ of~$\mathcal{Y}$. 
Combining the inverse of this bijection with~(\ref{Eq:RightMARIdentityII}) gives
\begin{equation}
   f(\mathbf{y}^{mt(\mathbf{r})} |\, \mathbf{y}^{ot(\mathbf{r})}) \,=\,
      f(\mathbf{y}^{mi(\mathbf{r})} |\, \mathbf{y}^{ob(\mathbf{r})}) \,=\,
      p(\mathbf{y}^{mi(\mathbf{r})} |\, \mathbf{y}^{ob(\mathbf{r})}, \mathbf{r}){,}
   \label{Eq:OtherSideOfProjection}
\end{equation}
{where the first equality is for values of functions with different domains.}
For notational simplicity, we relabel the {missingness} patterns, if necessary, so that $\mathbf{r} = \mathbf{r}_k$.
Conditioning on the $\mathbf{y}^{ot(\mathbf{r})}$ variables in (\ref{Eq:MixedUpPatterns}) yields
\begin{equation}
   f(\mathbf{y}^{mt(\mathbf{r})}|\, \mathbf{y}^{ot(\mathbf{r})}) \,=\,
      p(\mathbf{r})\,
      p(\mathbf{y}^{mi(\mathbf{r})} |\, \mathbf{y}^{ob(\mathbf{r})}, \mathbf{r}) \,+\,
      \sum_{j=1}^{k-1}
      p(\mathbf{r}_j)\,
      p(\mathbf{y}^{mt(\mathbf{r})} |\, \mathbf{y}^{ot(\mathbf{r})}, \mathbf{r}_j \,).
   \label{Eq:ConditionedMixedUpPatterns}
\end{equation}
Substituting (\ref{Eq:OtherSideOfProjection}) into (\ref{Eq:ConditionedMixedUpPatterns}) and rearranging then gives:
\begin{equation}
      p(\mathbf{y}^{mi(\mathbf{r})} |\, \mathbf{y}^{ob(\mathbf{r})}, \mathbf{r})  \,=\,
      \frac{1}{1 - p(\mathbf{r})}
      \sum_{j=1}^{k-1}
      p(\mathbf{r}_j)\,
      p(\mathbf{y}^{mt(\mathbf{r})} |\, \mathbf{y}^{ot(\mathbf{r})}, \mathbf{r}_j \,).
   \label{Eq:MarginalRemoved}
\end{equation}

When the data comprise $n$ IID draws with differing {missingness} patterns across units, holding $\mathbf{r}$ fixed
in~(\ref{Eq:MarginalRemoved}) and letting $\mathbf{y}$ vary shows that associations on the left hand side for which
data are never observed are partially observed on the right hand side amongst units with {missingness} patterns different
from~$\mathbf{r}$.
This key feature of MAR is obscured in the notation on the right hand side of~(\ref{Eq:WrongMARIdentity}).

% ==========================================================================

\end{document}